\documentclass[preprint]{aastex62}

\usepackage{graphicx}
\usepackage{hyperref}
\usepackage{amsmath}
\usepackage{natbib}
\usepackage{float}
\usepackage{soul}
\usepackage{ulem}

\shorttitle{Solar disk center shows scattering polarization in the Sr~{\sc i} 4607~\AA \ line}

\shortauthors{Zeuner et al.}
\begin{document}
	
	\title{\Large Solar disk center shows scattering polarization in the Sr~{\sc i} 4607~\AA \ line}
	
	%\received{January 22 2020}
	%\revised{February 13 2020}
	\accepted{April 4, 2020}
	
	\submitjournal{ApJ Letters}
	
	%\correspondingauthor{Franziska Zeuner}
	%\email{zeuner@irsol.ch}
	
	\author[0000-0002-3594-2247]{Franziska Zeuner}
	\affil{ Max-Planck-Institut f{\"u}r Sonnensystemforschung, Justus-von-Liebig-Weg 3, D-37077 G{\"o}t\-tingen, Germany}
	\affil{ Georg-August-Universit{\"a}t G{\"o}ttingen, Institut f{\"u}r Astrophysik, Friedrich-Hund-Platz 1, D-37077 G{\"o}ttingen, Germany}
	\affil{ Istituto Ricerche Solari Locarno (IRSOL, associated with USI), 6605 Locarno-Monti, Switzerland}
	
	\author{Rafael Manso Sainz}
	\affil{ Max-Planck-Institut f{\"u}r Sonnensystemforschung, Justus-von-Liebig-Weg 3, D-37077 G{\"o}t\-tingen, Germany}
	
	\author{Alex Feller}
	\affil{ Max-Planck-Institut f{\"u}r Sonnensystemforschung, Justus-von-Liebig-Weg 3, D-37077 G{\"o}t\-tingen, Germany}
	
	\author{Michiel van Noort}
	\affil{ Max-Planck-Institut f{\"u}r Sonnensystemforschung, Justus-von-Liebig-Weg 3, D-37077 G{\"o}t\-tingen, Germany}
	
	\author[0000-0002-3418-8449]{Sami K. Solanki}
	\affil{ Max-Planck-Institut f{\"u}r Sonnensystemforschung, Justus-von-Liebig-Weg 3, D-37077 G{\"o}t\-tingen, Germany}
	\affil{ School of Space Research, Kyung Hee University, Yongin, Gyeonggi-Do, 446-701, Republic of Korea}
	
	\author[0000-0003-1409-1145]{Francisco A. Iglesias}
	\affil{ Universidad Tecnol\'{o}gica Nacional - Facultad Regional Mendoza,  CEDS, Rodriguez 243, Ciudad, 5500, Mendoza,  Argentina}
	\affil{ Consejo Nacional de Investigaciones Cient\'{i}ficas y T\'ecnicas (CONICET), Godoy Cruz 2290, C1425FQB, Buenos Aires, Argentina}
	
	\author[0000-0001-8016-0001]{Kevin Reardon}
	\affil{ National Solar Observatory, 3665 Discovery Drive, Boulder, CO 80303, USA}
	
	\author[0000-0001-7764-6895]{Valent\'{i}n Mart\'{i}nez Pillet}
	\affil{ National Solar Observatory, 3665 Discovery Drive, Boulder, CO 80303, USA}

	\begin{abstract} 
		Magnetic fields in turbulent, convective high-$\beta$ plasma naturally develop highly tangled and complex topologies---the solar photosphere being the paradigmatic example. These fields are mostly undetectable by standard diagnostic techniques with finite spatio-temporal resolution due to cancellations of Zeeman polarization signals. Observations of resonance scattering polarization have been considered to overcome these problems. But up to now, observations of scattering polarization lack the necessary combination of high sensitivity and high spatial resolution in order to directly infer the turbulent magnetic structure at the resolution limit of solar telescopes. Here, we report the detection of clear spatial structuring of scattering polarization in a magnetically quiet solar region at disk center in the Sr~{\sc i} 4607~\AA~spectral line on granular scales, confirming theoretical expectations. We find that the linear polarization presents a strong spatial correlation with the local quadrupole of the radiation field. The result indicates that polarization survives the dynamic and turbulent magnetic environment of the middle photosphere and is thereby usable for spatially resolved Hanle observations. This is an important step toward the long-sought goal of directly observing turbulent solar magnetic fields at the resolution limit and investigating their spatial structure.
	\end{abstract}
	
	
	%-------------------------------------------------------------------
	\section{Introduction} 
	\label{sec:intro}
	%-------------------------------------------------------------------
	
	The observation of magnetic fields on the solar surface is
	crucial for the empirical study of fundamental astrophysical processes such as magnetoconvection, turbulent dynamos, magnetic reconnection, or plasma energization. The Zeeman effect, the splitting of spectral lines in the presence of a magnetic field and its associated polarized thermal emission, provides a direct spectroscopic technique to detect and diagnose magnetic fields in the solar atmosphere. However, there are fundamental limits for all Zeeman diagnostic techniques to detect magnetic fields that are highly tangled or turbulent and dynamic, because Zeeman polarization patterns from opposite polarities or crossed fields within the observa- tional resolution element cancel out. This makes magnetic fields undetectable in a regime of particular importance, for example, in connection with turbulent small-scale dynamo action and current dissipation. \par
	The Hanle effect, the magnetic modulation of resonance scattering polarization, does not suffer from such fundamental limitations \citep{Stenflo1994, LandiDeglInnocenti2004}.
	It creates a polarization pattern that depends on the scattering geometry and the magnetic field, even in the case of zero magnetic flux within the resolution element. \par 
	
	The discovery of the linearly polarized component of the spectrum observed close to the solar limb due to scattering was a breakthrough in solar physics that opened the possibility for diagnosing photospheric turbulent magnetic field configura- tions even at spatial scales below the resolution element 
	\citep{Stenflo1982,Faurobert-Scholl1993,TrujilloBueno2003,MansoSainz2004,TrujilloBueno2007,Stenflo1996}. 
	The Sr~{\sc i} line at 4607~\AA, one of the strongest scattering polarized lines  in the visible, has been widely studied. Its interpretation revealed an ubiquitous hidden magnetic field, which has been proposed to contribute significantly to the energy balance of the Sun \citep{TrujilloBueno2004}. \par
	
	To date, observations of scattering polarization on the photosphere have been done mostly close to the solar limb and lack spatial and temporal resolution. In practice, this means that we can neither directly probe magnetic fields nor track dynamo action at the smallest scales. Previous analyses of Hanle signals relied heavily on numerical modeling of the radiation field and required assumptions on the statistical properties of the magnetic field (its probability density function and spatial distribution), both remain largely unconstrained by the low resolution observations. Also, this makes it difficult to create a coherent picture of the magnetic field in the quiet Sun from the complementary Zeeman and Hanle observations. \par
	
	There is a growing interest in observing scattering polarization at lower heliocentric distances \citep{Malherbe2007, Bianda2011, Bianda2018, Zeuner2018, Dhara2019} that resolve the polarization fluctuations resulting from the local symmetry breaking of the radiation field with the aim of actually diagnosing the statistical distribution of magnetic fields at the smallest scales. This has been largely motivated by numerical radiative transfer calculations of scattering line polarization in realistic 3D models of the solar atmosphere, showing that horizontal inhomogeneities may indeed produce measurable linear polarization patterns at granular scales \citep{TrujilloBueno2007, DelPinoAleman2018}. This is of diagnostic value for small-scale dynamos and magnetoconvection \citep[e.g.,][]{Vogler2007,Rempel2014}.
	Observing this is a challenge because the polarimetric signals are weak and the high spatio-temporal resolution required severely limits the number of available photons in a way that cannot be compensated easily through standard techniques. Unlike intensity, polarization is quantified through signed quantities and simple averaging (either by low spatial resolution or post-processing) over areas characterized by very inhomogeneous radiation fields may lead to cancellations rather than signal enhancement. Therefore, observations have to provide at least subgranular resolution.\par
	
	This Letter reports the detection of spatially structured scattering polarization signals in the Sr~{\sc i} 4607~\AA \ line in a quiet Sun region at disk center. 
	It relies on a novel analysis technique that enhances weak scattering polarimetric signals using a reconstruction technique of the pattern of the local quadrupole moment of the radiation field inferred from the observed intensity map itself. We show that a reconstruction based on geometric considerations is sufficient for this purpose; therefore, our work is independent of models that include either magnetohydrodynamic or detailed radiative transfer calculations.
	
	%-------------------------------------------------------------------
	\section{Observational data} 
	\label{sec:obs}
	%-------------------------------------------------------------------
	
\begin{figure}[ht]
		\centering
		\includegraphics[width=0.7\textwidth]{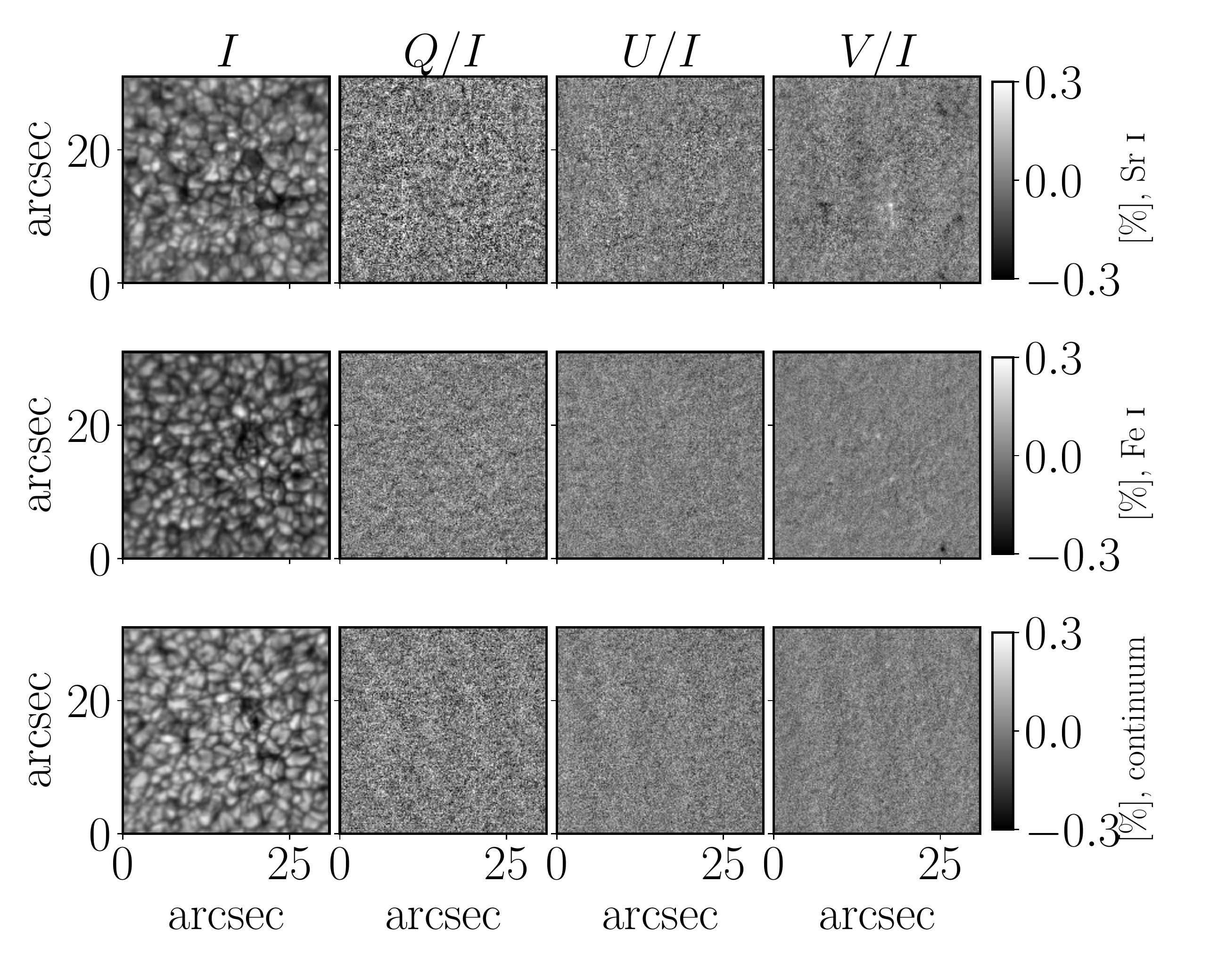}
		\caption{
			Time-averaged data set (over the entire observation period), consisting of the complete Stokes vector (columns) at the three different recorded wavelength positions (rows).
			The gray scale refers only to the normalized polarization Stokes images that have an rms noise level on the averaged images of $\sim$0.03\%. The root-mean-square (RMS) contrast for the Stokes $I$ images is $\sim6\%$.
			Granulation is clearly distinguishable in the temporally averaged images, showing the high quality of the data and the uninterrupted stable seeing conditions during the full observation. The seeing conditions in all wavelength positions were almost identical, allowing a fair comparison between the different data sets. Note that the $Q/I$ and $U/I$ Sr~{\sc i} core images are practically free of residual artifacts such as low-amplitude fringe patterns, as the data reduction steps are optimized for this wavelength position and are then applied to the data of the other two positions. Localized patches of both $V/I$ polarities in the core of the Sr~{\sc i} and Fe~{\sc i} lines are visible and of solar origin (longitudinal Zeeman signatures).} 
		\label{fig:stokesimages}
	\end{figure}

	The observations were carried out with our new high-cadence Fast~Solar~Polarimeter~2 at the Dunn Solar Telescope (see Appendix \ref{sec:ap_obs} for more details on the instrument and observation), which provides increased polarimetric sensitivity (\mbox{$<$ 0.1\%}) while conserving sufficient spatio-temporal resolution to 
	resolve subgranular scales (see Figure~\ref{fig:stokesimages}). We successively observed in three wavelength positions: the Sr~{\sc i} line core, the core of a neighboring Fe~{\sc i} line and a continuum position. The integration time at the former two positions was 210~s, while in the continuum it was 120~s. 
	The two latter positions serve as references. The filter was spectrally broad enough to average substantially over the line profiles (see Appendix \ref{sec:ap_obs} for the spectral filter profile). 
	After data reduction (see Appendix \ref{sec:ap_dare} for the data reduction steps), we have 42 (24) $\times$ 5~s averaged images per Stokes parameter 
	in the line cores (continuum). 
	We may safely assume that we capture an instantaneous solar scene in each 5~s image, compared to typical solar evolution times of around 30~s in the quiet Sun at 
	the spatial scales of our observation. 
	The observed region was located close to solar disk center ($\mu$=cos($\theta$)=0.98, $\theta$ is the heliocentric angle) and corresponded to the quiet Sun internetwork. 
	We checked the quietness of the region in H-$\alpha$ before we carried out the observation.

	%-------------------------------------------------------------------
	\section{Pixel classification based on the local quadrupole of the radiation} 
	\label{sec:method}
	%-------------------------------------------------------------------
	
\begin{figure}[ht]
		\centering
		\includegraphics[width=0.4\textwidth]{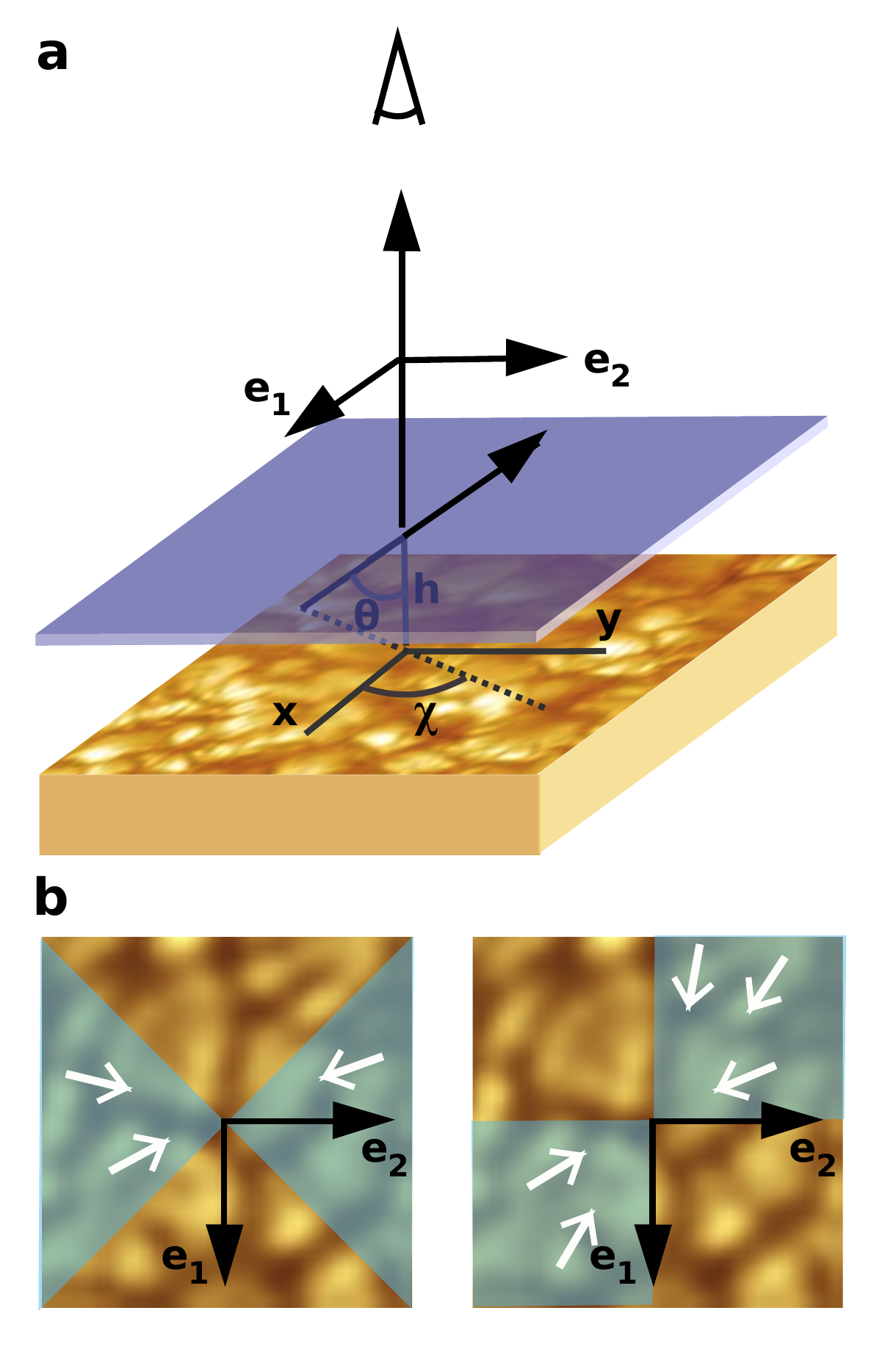} 
		\caption{Scattering polarization at the center of the solar disk arises from the axial symmetry breaking of the radiation field produced by inhomogeneities in the atmosphere.
			a) For a given point in the field of view, we use the observed intensity map dominated by the continuum and choose a reference system 
			with the $x$-axis along the positive-$Q$ direction ($\boldsymbol{e}_1$) and the $z$-axis along the line of sight. The $Q$ direction is given by the polarimeter used for the observation.
			The radiation, which hits a scatterer in an elevated thin and uniform layer (schematized by the blue surface) at a height of $h$ above the observed intensity map, is represented by the complex quadrupolar
			component of the radiation field ($J^2_2$). To estimate $J^2_2$ with equation~\ref{eq:j22}, we use the observed intensity map as a proxy of the  incoming radiation field below the scattering layer. Each ray of this radiation field hitting a scatterer can be characterized by the height $h$ and the azimuthal and polar angles $\chi$ and $\theta$, respectively. See the main text for details. b) As seen by the observer along the scattered direction,
			radiation incoming from the areas marked with white arrows (e.g., east-west for Stokes $Q$) contributes to positive $Q$
			($\operatorname{Re}\left(J^2_2\right)\equiv \tilde{J}^2_2<0$; left panel),
			or positive $U$ ($\operatorname{Im}\left(J^2_2\right)\equiv\hat{J}^2_2<0$; right panel) signals.
		}
		\label{fig:thefig}
	\end{figure}

	The scattering geometry for an observation at disk center is presented in Figure~\ref{fig:thefig}.
	The radiation scattered in the Sr~{\sc i} line is always linearly polarized, when incoming and scattered beams are perpendicular to each other.  
	Therefore, incoming radiation from east and west is scattered 
	toward the observer with $Q>0$,
	and radiation from north and south, with $Q<0$ ---similarly for $U$, but for the direction of the incoming radiation turned by 45$^{\circ}$ (panel (b) of Figure~\ref{fig:thefig}). Net linear polarization is therefore a result of an axially asymmetric illumination, e.g., $Q>0$ if the east-west component of the radiation field dominates.
	More quantitatively, the combination of the Stokes parameters $Q+{\rm i}\, U$ 
	of the scattered radiation is proportional to the quadrupolar component, $-\sqrt{3}\,J^2_2$, 
	of the incoming radiation field 
	\citep[see Figure~\ref{fig:thefig}; ][]{Chandrasekhar1960, LandiDeglInnocenti2004}. 
	
	We use a simple scattering model to estimate 
	the real and imaginary components of $J^2_2=\tilde{J}^2_2+{\rm i}\,\hat{J}^2_2$ 
	at any point in the field of view and at every time step from the respective observed intensity map. Note that due to the broad pre-filter the observed intensity maps, even at the line core wavelength positions, are dominated by the continuum intensity. The scattering model is simplified by using the following assumption: that a thin scattering layer is at a height $h$ above an atmosphere that provides incoming radiation given exactly by the observed intensity map. Thereby, we neglect the three-dimensional structure of the Sun's surface. In the following we will describe the procedure in more detail.
	
	The complex quadrupolar component of the radiation field characterizes the incoming radiation on a scatterer \citep{LandiDeglInnocenti2004}:
	\begin{equation}
	\label{eq:j22}
	J^2_2=\frac{\sqrt{3}}{4}\oint \frac{\mathrm{d}\Omega}{4\pi} \sin^2\theta\, {\rm e}^{-2{\rm i}\chi} I(\theta, \chi).
	\end{equation}
	
	The integral is over the unit sphere $\mathrm{d}\Omega=\mathrm{d}\mu\,\mathrm{d}\chi=\cos\theta \mathrm{d}\theta \mathrm{d}\chi$ with the polar and azimuthal components $\theta$ and $\chi$, respectively (for $\theta=0$, the incoming radiation is vertical, see Figure~\ref{fig:thefig} for the geometry). Only the intensity $I$ is relevant
	here, additional contributions from polarization are negligible. 
	The component $J^2_2$ may be estimated by considering a thin, uniform scattering layer model, at a height $h$ above an 
	atmosphere that radiates isotropically\footnote{Consequentially of this assumption, the center-to-limb variation of the underlying radiation field is neglected.} and radiates exactly like the observed intensity map. 
	The intensity $I(\theta, \chi)$ incoming on the scatterer at $(0, 0, h)$ (in our model $h$ is in units of a pixel) from a ray along 
	the direction $\theta$ and $\chi$, is then exactly given by the observed intensity at the point $-h\tan\theta(\cos\chi, \sin\chi, 0)$ 
	where the ray path intersects the underlying atmosphere. This is the only  point where the parameter $h$ enters the equation~\ref{eq:j22}. 
	Numerical integration over the outgoing hemisphere provides $J^2_2$ at any given point in the field of view (FoV). At the edge of the FoV we assume periodic boundary conditions.\par
	
	The value of the single parameter in the model is set to $h=3$~pixels, because  this is where the $J^2_2$ maps show the  greatest contrasts (see Appendix~\ref{sec:j22}).
	Regions with similar illumination (similar $J^2_2$) are expected to produce similar values of $Q$ and $U$, up to some factor times the random noise.\par
	In order to increase the polarization signal while decreasing the noise level, we follow a two-step process. First, we calculate $\tilde{J}^2_2$ and $\hat{J}^2_2$ based on the 5~s intensity images for each pixel in the FoV as explained in Figure~\ref{fig:thefig} and above. As we are not interested in the absolute values of $J^2_2$, we normalized $J^2_2$ according to the maximum value occurring 
	during the complete time series. Therefore, the $J^2_2$ values lie between $\pm 1$.
	
	Second, we temporally and spatially average the observed polarization only among pixels with similar values of $J^2_2$ (i.e., within 0.05 in the [-1,1] range).
	This process circumvents spatial smearing of the linear polarization signals during solar evolution while decreasing the noise level significantly.

	%-------------------------------------------------------------------
	\section{Results} 
	\label{sec:results}
	%-------------------------------------------------------------------
	
	According to the above picture, a scattering dominated 
	$Q$ signal should anticorrelate strongly with $\tilde{J}^2_2$ 
	(the larger the illumination from east-west with respect to 
	north-south, the smaller (more negative) $\tilde{J}^2_2$ and the larger the value of $Q$). Following a similar argument, $U$ should anticorrelate with $\hat{J}^2_2$.
	In the presence of a magnetic field, these correlations change in general.
	A magnetic field along (or away from) the line of sight in the Hanle regime, for example, rotates theses patterns so that for increasingly stronger fields, $Q$ and $U$ tend to 
	anticorrelate with $\hat{J}^2_2$ and $\tilde{J}^2_2$, respectively.
	However, for an isotropic distribution of magnetic fields, we recover the behavior 
	of the pure scattering case. The difference between the isotropic field and the pure scattering case lies in decreased amplitudes of $Q$ and $U$ compared to the zero-field case.
	This is exactly what we find.\par
\begin{figure}[ht]
		\centering
		\includegraphics[width=0.7\textwidth]{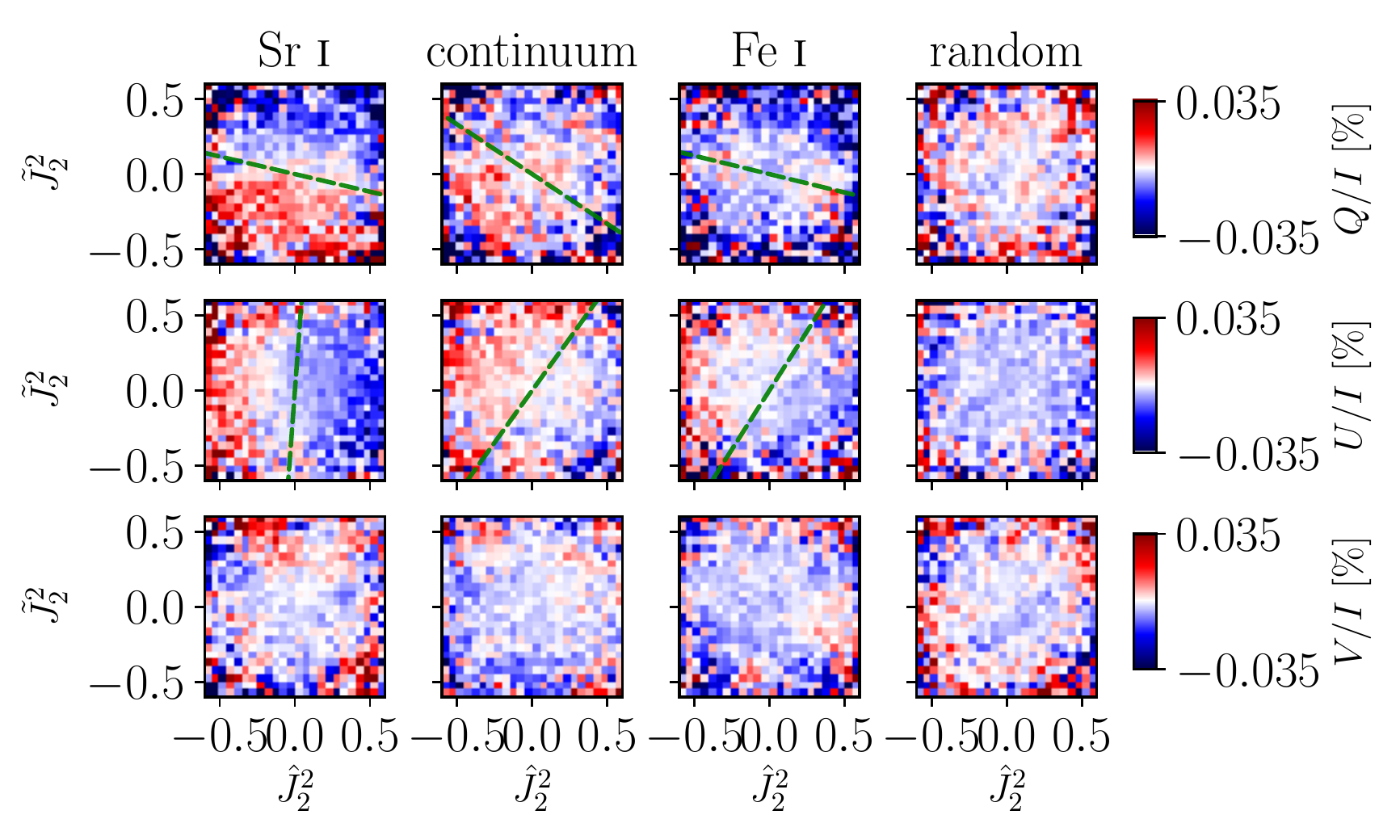} 
		\caption{
			Average (as explained in the section~\ref{sec:method}) observed polarization as a function of the normalized radiation's quadrupolar component $J^2_2$ (see the main text for explanation) for all three observed wavelength positions and for randomized Sr~{\sc i} polarization images (from left to right). 
			Here, we display only the polarization in the limited range of $-0.55 \leq J^2_2 \leq 0.55$, where most pixels lie.
			Green dashed lines show the nodal lines, which we get from a weighted two-dimensional 
			linear regression fit (see the main text for details).                     
			For the regression, $Q/I$ and $U/I$ were fit independently and their nodal lines (green dashed lines) appear appear roughly orthogonal and
			barely rotated in the $\tilde{J}^2_2$-$\hat{J}^2_2$ plane within observational uncertainties.
		}
		\label{fig:j_sum}
	\end{figure}

	Figure~\ref{fig:j_sum} shows the $Q/I$ and $U/I$ 
	in the $\hat{J}^2_2$-$\tilde{J}^2_2$ plane, when temporally and spatially averaged as explained in the last section. 
	The core of the Sr~{\sc i}~4607~\AA \ line clearly shows the anticorrelation between $Q/I$ and $\tilde{J}^2_2$, as well as between $U/I$ and $\hat{J}^2_2$.
	Such a pattern is barely noticeable in the continuum, the core of the Fe~{\sc i} line, 
	or net circular polarization ($V/I$). 
	We tested that the pattern in the Sr~{\sc i} line is not an artifact 
	of the analysis technique by repeating exactly the same process but randomly scrambling the positions of
	the pixels in the observed polarization images. 
	In this case, the coherence between polarization and radiation field structure disappears (rightmost panels in Figure~\ref{fig:j_sum}).

	The residual $U/I$ feature in the reference wavelength positions (Fe~{\sc i} core and continuum) is most likely an artifact due to the broad pre-filter in the observation setup, which in combination with the expanded tails of the spectral profile of the Fabry-P\'{e}rot system spectrally contaminates the reference wavelengths with Sr~{\sc i} signals. Stokes $Q/I$ in the reference wavelength position is probably missing an obvious residual feature like in Stokes $U/I$, as the noise level is higher in this Stokes parameter compared to $U/I$ (see Appendix~\ref{sec:ap_dare} for an explanation).
	We exclude the possibility that the features arise as residual Stokes $I$ cross-talk
	or the transverse Zeeman effect, since the reference wavelength positions should otherwise also show them as strongly as in the Sr~{\sc i}.
	Hence, we are confident that the linear polarization signals obtained after averaging are 
	actual scattering polarization signals 
	in the core of the Sr~{\sc i} line at the center of the solar disk. In addition, Figure~\ref{fig:j_sum} provides evidence that the polarization signal is structured with respect to the underlying radiation field.\par
	
	Here we do not try a detailed spectropolarimetric 
	modeling of the observations to infer the properties on the observed region;
	rather, we consider the qualitative, more fundamental implications of this detection: that the atomic alignment induced in the excited level, ${}^1P^\circ$, of Sr~{\sc i} by anisotropic scattering is not destroyed by depolarizing effects (either from collisions or magnetic fields), 
	even in the deep, dense regions of the photosphere probed at disk center. \par
	
	Vertical magnetic fields are very efficient at depolarizing and rotating the plane of scattering polarization
	from resolved horizontal inhomogeneities observed at disk center \citep{MansoSainz2011}. 
	
	To find the rotation angle of the scattering polarization plane with respect to the radiation field, we fit the plane $L=\alpha\cdot \hat{J}^2_2 +\beta\cdot \tilde{J}^2_2 $ to the polarization values in each panel in Figure~\ref{fig:j_sum} independently, without considering an offset, as the mean polarization was subtracted in the data reduction. The nodal lines, i.e., where $L=0$, indicate where the observed linear polarization is close to zero. In the Sr~{\sc i} line core, the $L=0$ lines are nearly horizontal, and almost perpendicular to each other for Stokes $Q/I$ and $U/I$, respectively.
	We do not find evidence that the small rotation in Figure~\ref{fig:j_sum} in the Sr~{\sc i} data, indicated by a mismatch of nodal lines $Q/I=0$ and $U/I=0$ with $\hat{J}_2^2=0$ and $\tilde{J}^2_2=0$, respectively, has any other source than noise. We tested this by adding a few realizations of random noise (with the same RMS as the observation itself) to the observation and repeat the fitting process for each realization. We find an angular uncertainty of about 5$^{\circ}$. \par
	Our findings (scattering polarization but no significant Hanle rotation) advocate two possible scenarios.
	If the photosphere is dominated by vertical magnetic fields, most of them must be well below the saturation regime---otherwise, the polarimetric signal would vanish completely,
	while from the lack of significant rotation of the polarization plane we conclude that such a vertical distribution of fields has mixed polarities. 
	The RMS of the magnetic field strength $\Delta B_{\rm rms}$ over the observed field of view is such that $2(3\Delta B_{\rm rms}/22.8$~G$)\lesssim 10$ in the collisionless limit 
	(here, the critical field 22.8~G appears from expressing the Larmor frequency in units of the Einstein coefficient for spontaneous emission $A_{u\ell}=2\times 10^8$~s$^{-1}$, a factor of two is related to the quadrupolar component of the radiation field and the majority of magnetic field strength values, assuming a Gaussian distribution, lies between $\pm3\Delta B_{\rm rms}$). 
	Collisions contribute additionally to relax atomic polarization and the critical field
	increases by a factor of $(1+\epsilon+\delta)$, where $\epsilon$ and $\delta$ are the rates of inelastic and elastic depolarizing collisions, respectively, normalized to $A_{u\ell}$ \citep[][]{LandiDeglInnocenti2004}.\par
	At typical formation heights of the Sr~{\sc i} line \citep[at around 170~km above optical depth unity, see ][]{DelPinoAleman2018}, $\epsilon$ is negligible \citep[e.g.,][]{Regemorter1962} while depolarizing collisions are significant \citep[e.g.,][]{DelPinoAleman2018}: $\delta\approx 4$ \citep{Faurobert-Scholl1995} --- $\delta\approx 2$ \citep{MansoSainz2014,DelPinoAleman2018}. Note that with increasing $\delta$, the scattering polarization generated is also reduced and if $\delta\gg 1$, 
	it would be completely wiped out. 
	Therefore, our observations are compatible with a photosphere filled with (nearly) vertical 
	magnetic fields with $\Delta B_{\rm rms}\lesssim (38$~G$)(1+\delta)\approx 114-190$~G.
	Alternatively, the magnetic field could adopt a more general distribution, inclined to the vertical and having uniformly random azimuths. Our observational setup does not allow us at present to constrain horizontal magnetic fields.
	The vertical component depolarizes the scattering signals but a net polarization still emerges from regions dominated by the horizontal magnetic fields.\par
	For example, an isotropic, Maxwellian distribution of magnetic field strengths (average field strength $\bar{B}$), 
	results in a decrease of observed scattering polarization signal
	(with respect to the maximal, zero-field value). The observed scattering polarization amplitude decreases with increasing $\bar{B}$ but saturates at 20\% of the original polarization for $\bar{B}\gtrsim 130$~G.
	
	\begin{figure}[ht]
		\centering
		\includegraphics[width=0.7\textwidth]{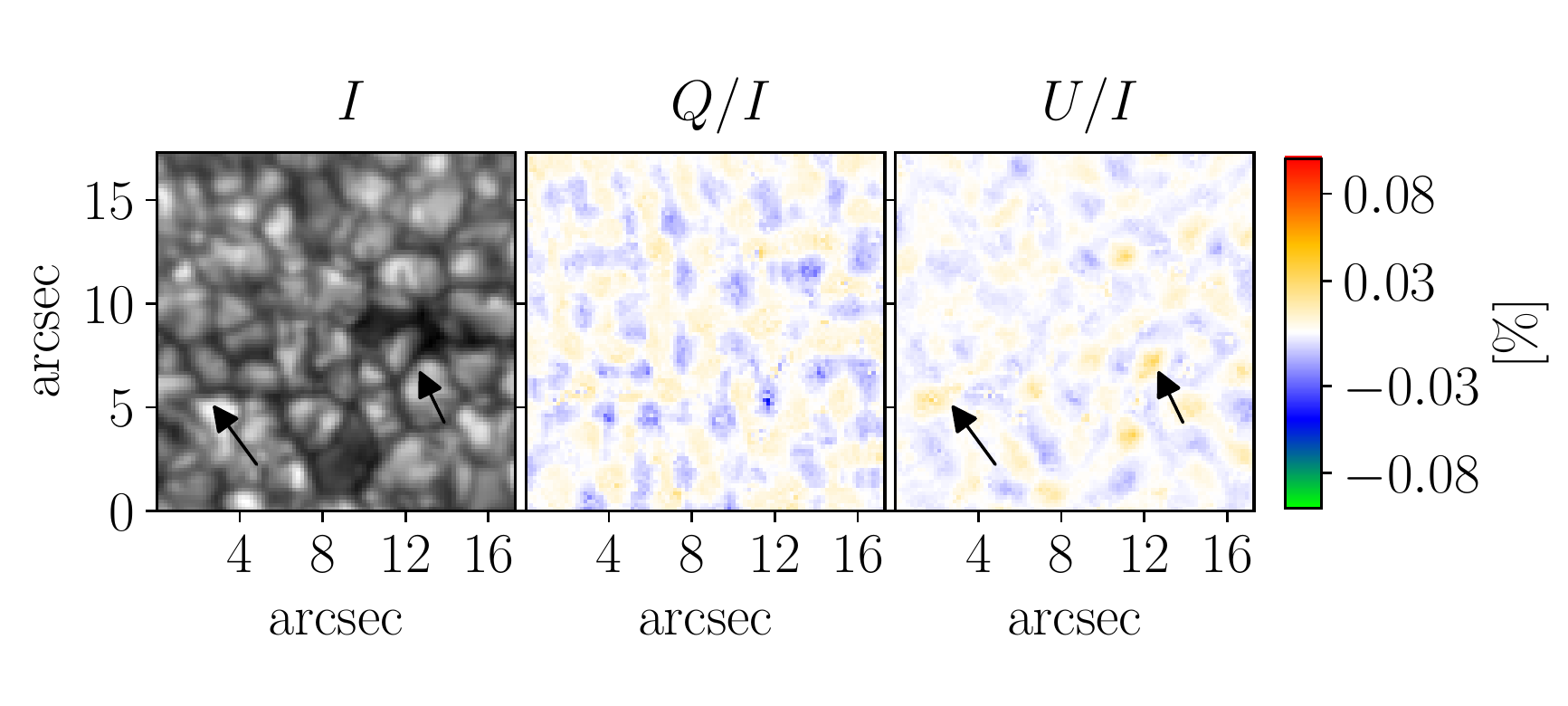} 
		\caption{Stokes $I$ and ``reconstructed" linear polarization maps in the Sr~{\sc i} line core.
			From left to right: temporally averaged observed intensity image (section of the top left frame of  Figure~\ref{fig:stokesimages}), reconstructed $Q/I$ and $U/I$ spatial maps from the polarization 
			in Figure~\ref{fig:j_sum}. These maps have been reconstructed by retracing the averaged polarization values from Figure~\ref{fig:j_sum} to their original positions
			in an observed 5~s Stokes $I$ image (i.e., to the places with the corresponding $J^2_2$ values) and then (again) temporally averaging over the full period of observation. The reconstruction can be considered to be equivalent to a clever way of denoising the observations.
		}
		\label{fig:newmap}
	\end{figure}

	Figure~\ref{fig:newmap} shows a reconstruction of the spatial distribution of the polarization signals.
	The reconstruction has been obtained by retracing the averaged polarization values shown in Figure~\ref{fig:j_sum} 
	to each spatial pixel according to its local value of the radiation field, i.e., for each 5~s $J^2_2$ map. Then we temporally averaged the reconstructed maps to achieve comparability to Figure~\ref{fig:stokesimages}.
	Note that multiple spatially different pixels in the reconstructed map may therefore have the same polarization values due to the prior averaging in Figure~\ref{fig:j_sum}.\\
	
	The spatial structure of the polarization signals is subgranular.
	They are mostly located at the interface between granules and intergranules. Notice that large signals with $U/I\sim0.05\%$ are visible and not limited to individual pixels, but extend to a size of about an arcsecond (marked with arrows in the $U/I$ reconstructed image and in the same spatial positions in the Stokes $I$ image to emphasize that the signals are located close to the interface between granules and intergranules). \par
	An upper limit of the average scattering structure size is $0.75\ensuremath{^{\prime\prime}}$ (which was found by convolving spatially with Gaussians of increasing widths until the resultant polarization amplitude in the $\tilde{J}^2_2$-$\hat{J}^2_2$ plane is in the order of the reference wavelengths). However, larger structures than $0.75\ensuremath{^{\prime\prime}}$ are occasionally possible, see Figure~\ref{fig:newmap}. These structure sizes are 
	consistent with the previous estimate of  $0.5\ensuremath{^{\prime\prime}}-1\ensuremath{^{\prime\prime}}$         
	\citep{Zeuner2018}. 
	
	The mean absolute linear polarization for the Sr~{\sc i} line core in Figure~\ref{fig:newmap} is 0.005\% and 0.003\% in Stokes $Q/I$ and $U/I$, respectively, while the standard deviation is 0.006\% and 0.005\%. These values obviously correspond to the finite spectral and spatial resolution of the observations and we expect higher signals with increased spatial and in particular, spectral resolution. 
	\citet{DelPinoAleman2018} report disk center polarization amplitudes of up to 1\% in synthetic Stokes profiles from recent magnetoconvection models of the quiet solar photosphere.
	\par

	%-------------------------------------------------------------------
	\section{Discussion and conclusion} 
	\label{sec:conclusion}
	%-------------------------------------------------------------------

	We have shown here how it is possible to map the scattering polarization component 
	of radiation at solar disk center by employing the information available on the inhomogeneous solar surface. With our observational setup we find absolute polarization amplitudes of 
	typically $\sim$0.004\%, corresponding to standard deviations of below $0.01\%$, while the mean spatial extent is limited to sizes smaller than 
	$0.75\ensuremath{^{\prime\prime}}$. It is known that the amplitude of the scattering polarization, in particular, is sensitive to the spectral resolution of the observation. 
	This has been studied in detail by \citet{DelPinoAleman2018}.
	In our case, the spectral resolution was about 67~m\AA \ FWHM and the polarization amplitude is even below the noise level of the observed polarization images, which gives a signal-to-noise ratio of less than one. This explains the absence of distinct polarization signals in the observed polarization images. The most striking result is that we find the scattering polarization to be anticorrelating with the radiation field's quadrupole tensor element $J^2_2$ estimated from the observed Stokes $I$ map, which is dominated by the continuum radiation in our data. We also find that the spatial structure of the scattering polarization is subgranular, in agreement with the theoretical predictions by \citet{TrujilloBueno2007} and \citet{DelPinoAleman2018}.
	This means that atomic polarization is not destroyed completely even in very dense layers of the quiet solar photosphere, confirming the results from numerical calculations reported by  \citet{TrujilloBueno2007,DelPinoAleman2018}.\par
	We find that the observed scattering polarization is compatible with two alternate scenarios of the magnetic field's structure in the quiet photosphere. Either the magnetic field in the photosphere is dominantly vertical, with strengths below the saturation regime of 114 - 190~G, depending on the collisional rates and of zero mean flux. Alternatively, the magnetic field can be much stronger if it is more horizontal on average. 
	These findings are not constrained by the specific spatial resolution of the observation, as the Hanle effect does not suffer from subresolution cancellations. However, resolving subgranular scales is necessary in order to resolve the polarization emerging from the axial symmetry breaking of the radiation field introduced by the thermal inhomogeneity due to granulation. Investigations based on the Hanle effect \citep{TrujilloBueno2004, DelPinoAleman2018} and
	Zeeman effect-based analyses \citep[e.g.][]{Lites2008,  Danilovic2010a, Danilovic2016} have returned average field strengths reaching
	100~G or more in the quiet Sun. Such significant average field strengths are consistent with our analysis, which are also found in recent magnetoconvection simulations with small-scale dynamo action \citep{Vogler2007, Schussler2008, Rempel2014,Khomenko2017}, particularly in the middle photosphere, where the Sr~{\sc i} line
	is formed. In the future, we expect stronger constraints by combining
	spatially resolved observations at different limb distances.
	
	The new polarimeter combined with the simple scheme used here to extract the signal from the noise was designed as a proof-of-principle experiment. Better results are expected from an improved observational setup, e.g, by utilizing a narrower filter to minimize the contamination of signals
	between lines and also the continuum; ideally, spectropolarimetry would avoid such contamination entirely.
	The novel technique employed here for the first time cleverly trades spatial information in the 
	whole FoV for polarimetric sensitivity. Therefore, there is much to gain from new 
	advanced instrumentation in upcoming facilities, such as the Daniel K. Inouye Solar Telescope currently being tested on Maui, Hawaii, that offer an extended 
	FoV, increased photon flux, and yet high spatial resolution. 
	Alternatively, extended time series, even of restricted FoV's,
	can now be integrated without loss of polarimetric signal.
	Once any of this is achieved and we are able to detect the 
	polarimetric signal and to quantify its statistics reliably,
	it will be possible to diagnose the magnetic field from the Hanle 
	signals solely from the statistics of the data in a totally model independent manner.
	Spatially resolved scattering polarimetry observation on the solar disk is one step closer to delivering on the promise of Hanle effect diagnostics in the photosphere that are complementary to, 
	but potentially on par with, state-of-the-art Zeeman techniques in spatial and temporal resolution.
	
	%-------------------------------------------------------------------
	\acknowledgments
	We thank the anonymous referee for a constructive report that has improved the quality and clarity of this work.
	The Fast~Solar~Polarimeter project is funded by the Max Planck Society (MPG) and by the European Commission, grant No. 312495 (SOLARNET). This project has received funding from the European Research Council (ERC) under the European Union's Horizon 2020 research and innovation programme (grant agreement No. 695075) and has been supported by the BK21 plus program through the National Research Foundation (NRF) funded by the Ministry of Education of Korea. The participation of F.Z. was funded by the International Max Planck Research School for Solar System Science. The National Solar Observatory (NSO) is operated by the Association of Universities for Research in Astronomy, Inc. (AURA), under cooperative agreement with the National Science Foundation.
	The authors declare that they have no competing financial interests.
	
	% %-------------------------------------------------------------------
	% %-------------------------------------------------------------------
	\appendix
	% %-------------------------------------------------------------------
	% %-------------------------------------------------------------------
	\section{Observation}
	\label{sec:ap_obs}
	% %-------------------------------------------------------------------
	
	All four Stokes parameters were recorded in a quiet Sun region very close to disk center ($\mu=0.98$) on 2017 August 8 between 14:41 and 15:20 UTC. The adaptive optics system was locked on the granulation. The observations were performed with the FSP~2, which is briefly described below. The polarimeter was integrated into a single-etalon collimated setup at the 0.76~m Dunn Solar Telescope (NSF's DST), located on Sacramento Peak in Sunspot, New Mexico.
	
	The Fast~Solar~Polarimeter (FSP) is a ground-based solar polarimeter designed to provide fast modulation and high frame rates to allow for suppression of seeing-induced cross-talk and image restoration at the same time. Additionally, high polarimetric accuracy is achieved. For a detailed description of the prototype of FSP, see \citet{Iglesias2016}. 
	In the second generation of FSP, used to obtain the data analyzed here, the pnCCD of the prototype of FSP was replaced by a 4k$\times$3k CMOS sensor. The frame rate was set to 200~Hz. To enable a dual-beam configuration on a single sensor, a custom designed\footnote{Max Planck Institute for Solar System Research in collaboration with LightMachinery Inc. The extinction ratio over the entire wavelength range of 400~nm - 860~nm is better than 1:40 in both channels.} polarizing beam-splitter was attached to the CMOS camera. The camera will be described in more detail below. The combined instrument, consisting of the modulator \citep{Iglesias2016}, the polarizing beam-splitter, and the camera is called FSP~2. \par
	
	The CMOS camera attached to the beam-splitter is a commercially available camera, based on the 
	CMOSIS CMV-12000 image sensor, which has been customized to include a sensor cold finger 
	combined with a thermo-electric temperature control, water cooling, and a sensor mask to shield 
	border pixels for calibration purposes. 
	
	The temperature  of the sensor is stabilized to 
	$20^{\circ}$~C to better than $\pm$0.5~K. The pixel area is 5.5~$\mu$m$^2$. The camera has a 
	readout noise of 13~e$^-$RMS. With a nonlinearity term of up to 2\%, the low contrast 
	images of the quiet Sun are affected by telescope induced polarization, that adds spurious 
	signals proportional to Stokes $I$ \citep{Keller1996} to the normalized polarization data. 
	Correction of the imprinted and scaled Stokes $I$ by simple subtraction results in a higher 
	noise level compared to data that are unaffected by telescope polarization. The scaling factor 
	for correcting this Stokes $I$ cross-talk is estimated and explained in more detail in Appendix \ref{sec:ap_dare}.
	The data depth is 10 bit. 
	The camera is directly attached to the beam-splitter, which prevents relative motion between these two components. Furthermore, the separation line of the channels in the dual-beam setup divides the camera sensor into two halves. To achieve the desired frame rate of 200~fps (modulation frequency was 50~Hz to obtain full Stokes measurements), the total readout area of the sensor was reduced to 2048$\times$2048 pixel$^2$. Consequently, when critically sampled at 4607~\AA \ (the plate scale is 0.062\ensuremath{^{\prime\prime}} pixel$^{-1}$), each dual-beam channel exhibits a field of view of 63\ensuremath{^{\prime\prime}} $\times$ 126\ensuremath{^{\prime\prime}}.
	
\begin{figure}[H]
		\centering
		\includegraphics[width=0.7\textwidth]{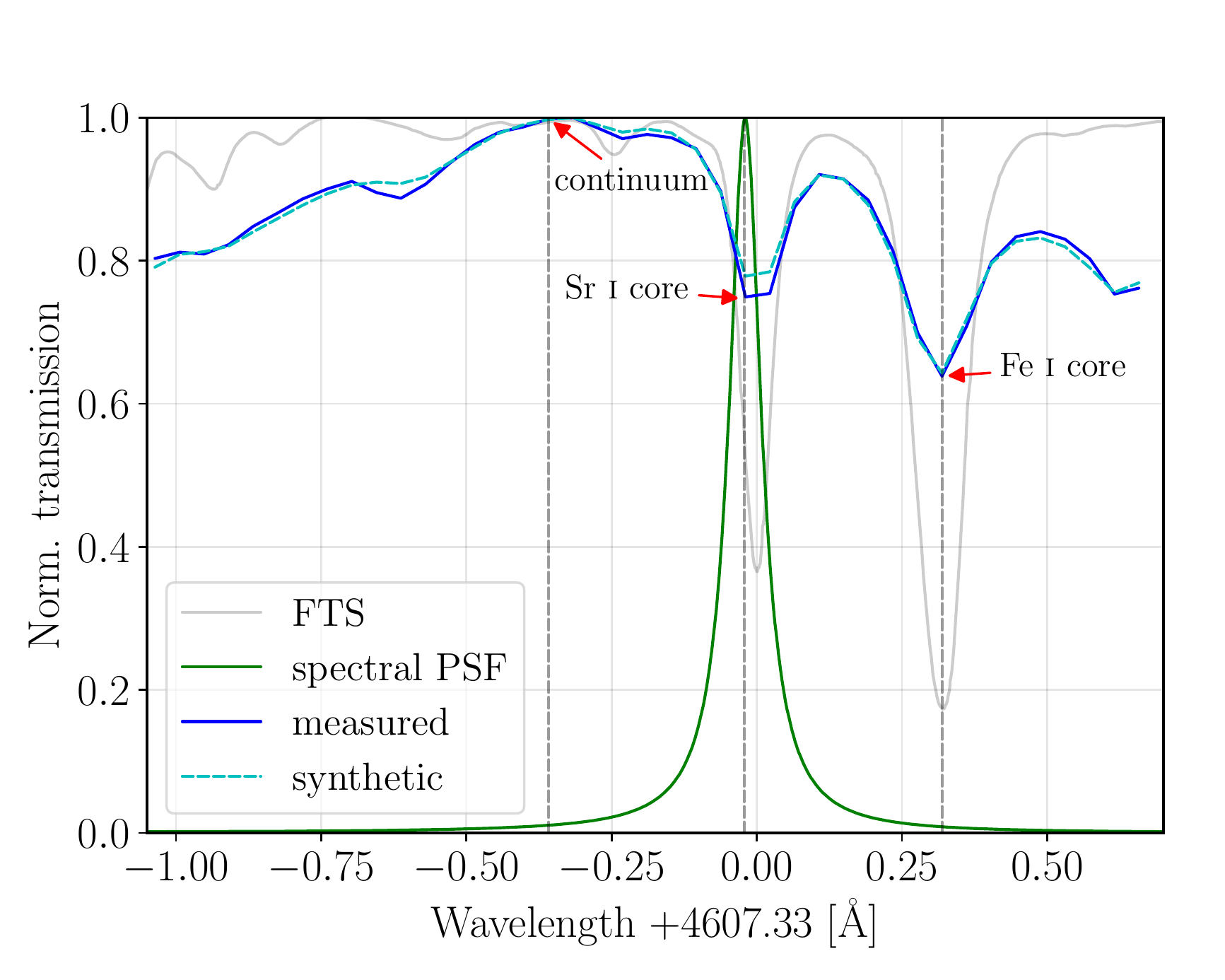}
		\caption{Spectral transmission profile (green line) of the filtergraph setup, a measured spectrum (blue) and synthetic spectrum obtained from a convolution of FTS data \citep{Neckel1999} with the spectral transmission profile. For the measured spectrum the Fabry-P\'{e}rot was tuned with a wavelength step of $\sim$8~m\AA. The spectral locations of the observed wavelength positions are marked with arrows.}
		\label{fig:spectrum}
	\end{figure}

	For wavelength discrimination, a single Fabry-P\'{e}rot etalon was used in a collimated setup with a spectral full-width at half-maximum (FWHM) of 67~m\AA. The free spectral range is 1.75~\AA. The pre-filter FWHM is 169.7~m\AA. This means, that secondary transmission peaks are insufficiently suppressed and spectral stray light affects the data, decreasing the SNR. The total normalized spectral profile for the Sr~{\sc i} line core position of the  Fabry-P\'{e}rot is shown in Figure~\ref{fig:spectrum} (secondary peaks are not visible in the displayed narrow frequency window). \par
	
	We sampled three wavelength positions around the Sr~{\sc i} line, which included a wavelength position very close to the Sr~{\sc i} line core, the neighboring Fe~{\sc i} line core, and a continuum point, displayed in Figure~\ref{fig:spectrum}. 
	
	Note that the sampled wavelength position was coincidentally shifted by about 20~m\AA \ into the blue wing of the Sr~{\sc i} line. Note also that we refer to a line core position, but the observation with the broad filter profile returns an integrated spectral line profile. In the Fe~{\sc i} line core position, the observed wavelength position coincides with the nominal line core position very well. At both spectral line core positions we exposed for 210~s, while at the continuum position we exposed for 120~s. 
	The data needed for calibration, i.e., polarization calibration, dark-field and disk center flat-field images are taken within one hour before and after the science data recording. 
	
	% %-------------------------------------------------------------------
	\section{Data reduction}
	\label{sec:ap_dare}
	% %-------------------------------------------------------------------
	
	The data reduction steps corrected for dark current, remove common mode and flat-field errors. A polarimetric demodulation is applied. Both channels of the beam-splitter are aligned and fringes are removed. One of the steps takes care of intensity to Stokes \{$Q$, $U$, $V$\} cross-talk as well as Stokes $V$ to \{$Q$, $U$\} cross-talk. The line core and the continuum images are also corrected. In the following, we will give an overview of these steps in the same order as they are applied. \par
	In the first step, a low-noise dark image is subtracted.
	Common mode errors are visible as offsets in pixels belonging to one sensor row due to electrical potential fluctuations in the readout hardware. For the common mode correction, in each frame the signal in 30 shielded pixels for each row on the left and right sides of the sensor are averaged. The average is subtracted from the respective row. \par
	Two hundred and fifty individual frames per modulation state are averaged, which corresponds to 5~s integration time. With a total of 3.5and 2~minute observation intervals in the spectral line cores and continuum, 42 and 24 modulation state images, respectively, are obtained per two-beam channel. 
	The flat-field is dark current corrected and common mode corrected. Then the individual images are flat-fielde Although possible, we decided not to apply multi-object, multi-frame blind deconvolution \citep[MOMFBD, see][]{vanNoort2005} restoration
	to the images to keep the noise distribution as close to the
	original as possible. The seeing was stable enough to provide high-resolution data without the restoration.
	
	To align the two channels of the dual-beam setup, we applied a mapping resulting from aligning reduced dot-target images. The numerical routine for alignment fragments the dot-target images and subaligns these fragments, while rotation, (de)magnification, and horizontal and vertical shifts are taken into account.
	
	To calibrate the data, we used known polarization states generated by the polarization calibration optics at the DST. The polarimetric efficiencies calculated from the modulation matrix after calibration are 0.42, 0.59 and 0.54 for $Q$, $U$ and $V$, respectively. 
	
	We applied a rotation matrix around the Poincar\'{e} sphere axes $U$ and $Q$ for a heuristic cross-talk correction between $V$ and \{$Q$, $U$\} in both channels separately, but field independent (i.e., all pixels in the FoV are rotated with the same angle). With this correction, we take care of residual telescope polarization.
	The angles of rotation are found by minimizing the cross-talk from the strongest Stokes $V$ signal to the linear polarization images. The estimated cross-talk for $V$ to $Q$ is corrected by 33$^{\circ}$ and 28$^{\circ}$ rotation around the $U$ axis for the each of the dual-beam channels. In an analogous manner, for a small cross-talk from $V$ to $U$, we corrected with a -3$^{\circ}$ and -8$^{\circ}$ rotation around the $Q$ axis. \par 
	
	Large-scale polarized fringes were visible in polarimetric Stokes $U$ and $V$ Sr~{\sc i} line core images. The polarized fringe removal is done by taking a 2D Fourier transform of the images. A ring-shaped mask is defined and, subsequently, the masked frequencies are removed. The mask is identical for the Stokes $U$ and $V$ images. The removed frequency components are replaced by interpolations from the remaining neighbor pixels and the image is inverse Fourier transformed.

	The polarimetric images are then normalized to Stokes $I$ at the corresponding wavelength. After normalization, the expected mean linear polarization signal is very low for an averaged quiet Sun region at disk center in the line core \citep[see][]{Neckel1999}. Therefore, we subtracted the spatial average of $Q$/$I$, $U$/$I$, and $V$/$I$ from the respective images to correct for residual instrumental polarization. Stokes $Q/I$ is the parameter which suffers the most from an offset of about $1\%$.
	
	In the last step, we corrected for cross-talk from Stokes $I$ due to sensor nonlinearity. For this correction, it is necessary to determine the factor $f$, followed by subtraction of $f\cdot I$ from the respective normalized polarimetric image. For each polarimetric Stokes \{$Q/I$, $U/I$, $V/I$\} image as well as for each time step  we determine $f$ separately. To find $f$, we minimized the rms of the image, as more structure from the Stokes $I$ cross-talk will result in a higher RMS value. The largest $f$ coefficients were found for the Stokes $Q/I$ images, which is not surprising, as here we found the largest spatially averaged polarimetric offset, indicating a large telescope polarization. Telescope polarization combined with sensor nonlinearity results in significant Stokes $I$ cross-talk. The $f$ values for $U/I$ were at least one order of magnitude smaller than for $Q/I$. Therefore, an increased noise level in Stokes $Q/I$ compared to $U/I$ is due to the ad hoc cross-talk correction from Stokes $I$. Finally, the images of both dual-beam channels are averaged and the mean values of each of the polarization Stokes images are subtracted.\par

	From an azimuthally averaged power spectrum, we estimate a resolution of 0.4\ensuremath{^{\prime\prime}}. Based on this estimated resolution we further spatially binned 3$\times$3 pixel$^2$ to a sampling of 0.19\ensuremath{^{\prime\prime}} pixel$^{-1}$.

	% %-------------------------------------------------------------------
	\section{$J^2_2$ maps}
	\label{sec:j22}
\begin{figure}[ht]
		\centering
		\includegraphics[width=0.7\textwidth]{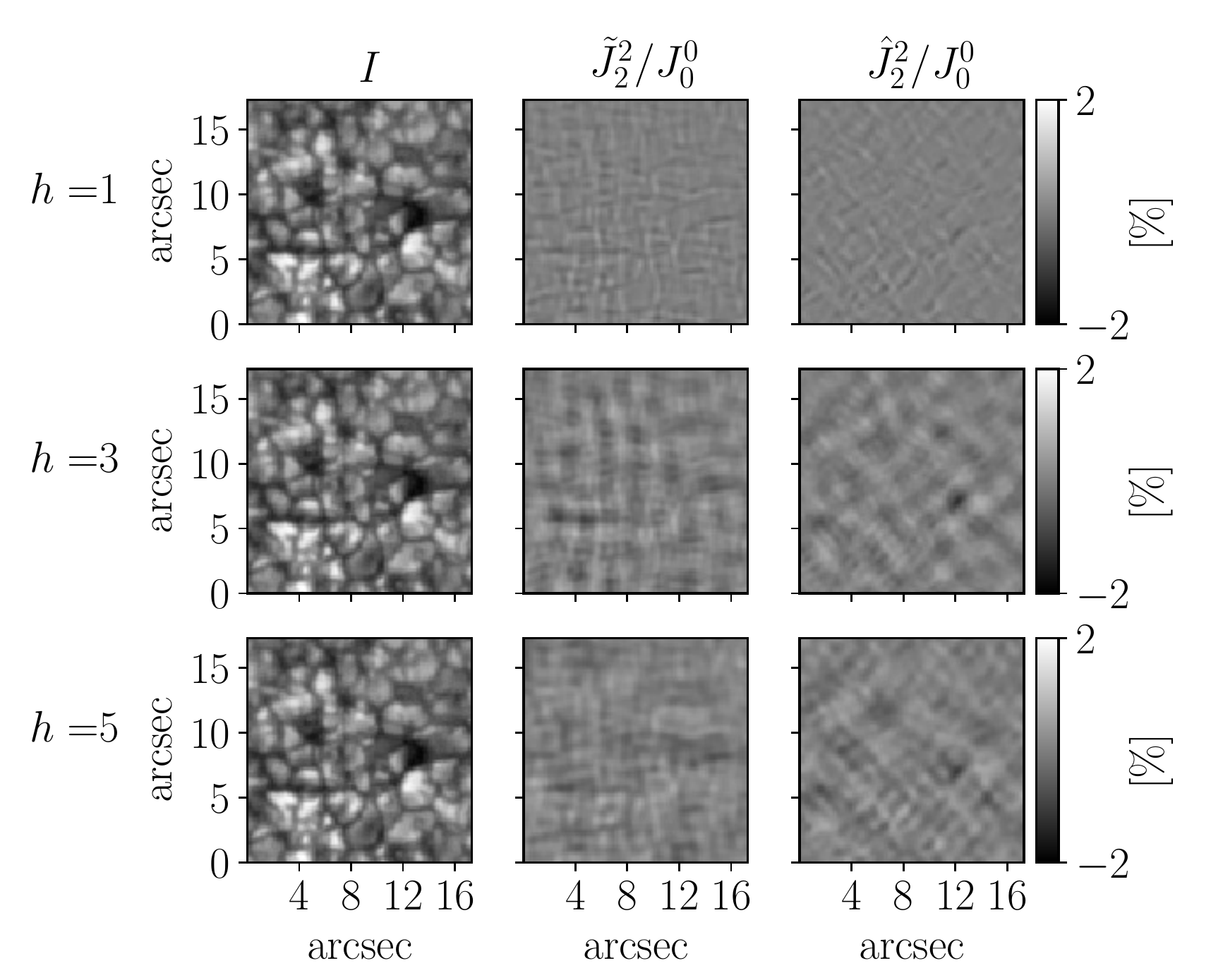} 
		\caption{Normalized real (center panels) and imaginary (right panels) parts of $J^2_2$, calculated from the observed (5~s) intensity Stokes $I$ images (left panels) with equation~\ref{eq:j22}. The gray scale on the right of the figure only applies to the normalized $J^2_2$ maps. The grey scale of the (to the maximum value) normalized Stokes $I$ images is between 0.65 and 0.98. From top to bottom the parameter $h$ (given in pixels) is increased from 1 to 5 in steps of 2, which has no effect on the Stokes $I$ image, but influences  $J^2_2$.}
		\label{fig:j22}
	\end{figure}

	% %-------------------------------------------------------------------
	Here we display the complex quadrupole moment of the radiation field $J^2_2$ in a hypothetical scattering layer estimated from a 5~s Stokes $I$ image in Figure~\ref{fig:j22}. Use was made of equation~\ref{eq:j22}, where the incoming intensity $I(\theta, \chi)$ on the scattering layer depends on the distance $h$ between the scattering layer and the atmosphere radiating at an intensity given by Stokes $I$.
	The only free parameter of the model to calculate the $J^2_2$ maps is therefore the height $h$. We varied the parameter $h$ between 1 and 5 spatial pixels, and found that the $J^2_2$ maps with $h=3$~pixels show the highest contrasts (about 10\% more than the $h$=5~pixel case). The contrast is a proxy for the spatial correlation of the radiation field, which, in turn, determines the structure sizes in the $J^2_2$ maps. The highest contrast is achieved where the $J^2_2$ structures are in the order of the correlation length of the underlying radiation field. \par
	We tested if $h=3$~pixels is the best choice by repeating the analysis for all three $h$ cases, and find that the reconstructed polarization signals are almost twice as large for the $h=3$~pixels case than for the other two cases.
	With our achieved pixel sampling, 3~pixels would correspond to a physical height of about 150~km above optical depth unity. This is a reasonable number, as \cite{Bommier2005a} found a formation height for Sr~{\sc i} between 220~km and 330~km at limb distances $\mu$=0.55 and $\mu$=0.09, respectively. Since we observed close to the disk center, we expect to sample deeper layers when observing in the Sr~{\sc i} core. Detailed theoretical investigations by \citet{DelPinoAleman2018} revealed at disk center an average height of formation of the Sr~{\sc i} line of 170~km above optical depth unity.
	
	%-------------------------------------------------------------------
	% References – no more than 50 references
	%------------------------------------------------------------------
	
	\bibliography{main.bib}
	\bibliographystyle{aasjournal}
	
\end{document}